\documentclass{article}
   \usepackage{graphicx}
   \usepackage{amssymb}
   \usepackage{epstopdf}
   \usepackage{apjgalley}
   \usepackage{multicol}
   \usepackage{psfig}
             \font\sevenrm=cmr7

\def\jetp{Sov. Phys. JETP}                      
\def\teq#1{$\, #1\,$}                           
\def\erg{\varepsilon}

\def\lambar{\lambda\llap {--}}
\def\fsc{\alpha_{\hbox{\sevenrm f}}}                                
\def\dover#1#2{\hbox{${{\displaystyle#1 \vphantom{(} }\over{
   \displaystyle #2 \vphantom{(} }}$}}

\def\Gammave{\overline{\Gamma}_{n0}}   
\def\Powerave{\overline{P}_{n0}}   
\def\today{\ifcase\month\or
  January\or February\or March\or April\or May\or June\or
  July\or August\or September\or October\or November\or
  December\fi
  \space\number\day, \number\year}
\begin{document}
%
%
\newcommand{\vol}[2]{$\,$\rm #1\rm , #2.}                 
\newcommand{\figureout}[5]{\centerline{}
   \centerline{\hskip #3in \psfig{figure=#1,width=#2in}}
   \vspace{#4in} \figcaption{#5} }
\newcommand{\twofigureout}[3]{\centerline{}
   \centerline{\psfig{figure=#1,width=3.4in}
        \hskip 0.5truein \psfig{figure=#2,width=3.4in}}
    \figcaption{#3} }    
\newcommand{\figureoutpdf}[5]{\centerline{}
   \centerline{\hspace{#3in} \includegraphics[width=#2truein]{#1}}
   \vspace{#4truein} \figcaption{#5} \centerline{} }
\newcommand{\twofigureoutpdf}[3]{\centerline{}
   \centerline{\includegraphics[width=3.4truein]{#1}
        \hspace{0.5truein} \includegraphics[width=3.4truein]{#2}}
        \vspace{-0.2truein}
    \figcaption{#3} }    

\newcommand{\figureoutfixed}[2]{\centerline{}
   \centerline{\psfig{figure=#1,width=5.5in}}
    \figcaption{#2}\clearpage }
\newcommand{\figureoutsmall}[2]{\centerline{\psfig{figure=#1,width=5.0in}}
    \figcaption{#2}\clearpage }
\newcommand{\figureoutvsmall}[2]{\centerline{\psfig{figure=#1,width=4.0in}}
    \figcaption{#2}\clearpage }
\newcommand{\tableout}[4]{\vskip 0.3truecm \centerline{\rm TABLE #1\rm}
   \vskip 0.2truecm\centerline{\rm #2\rm}   
   \vskip -0.3truecm  \begin{displaymath} #3 \end{displaymath} 
   \noindent \rm #4\rm\vskip 0.1truecm } 
%
%
%
%


\title{SPIN-DEPENDENT CYCLOTRON DECAY RATES IN STRONG MAGNETIC FIELDS}

   \author{Matthew G. Baring}
   \affil{Department of Physics and Astronomy MS-108, \\
      Rice University, P.O. Box 1892, Houston, TX 77251, U.S.A.\\
      \it baring@rice.edu\rm}

   \bigskip
    
   \author{Peter L. Gonthier}
   \affil{Hope College, Department of Physics and Engineering,
          27 Graves Place, Holland, MI 49422, U.S.A. \\
      \it gonthier@physics.hope.edu\rm}

    \and

   \author{Alice K. Harding}
   \affil{NASA's Goddard Space Flight Center, Laboratory for High Energy
      Astrophysics, Code 661 \\
      Greenbelt, MD 20771, U.S.A. \\
      \it harding@twinkie.gsfc.nasa.gov\rm}
\slugcomment{To appear in \it The Astrophysical Journal\rm , 
 Vol 630, September 1, 2005 issue.}
%

\begin{abstract}  
Cyclotron decay and absorption rates have been well studied in the
literature, focusing primarily on spectral, angular and polarization
dependence. Astrophysical applications usually do not require retention
of information on the electron spin state, and these are normally
averaged in obtaining the requisite rates.  In magnetic fields, higher
order quantum processes such as Compton scattering become resonant at
the cyclotron frequency and its harmonics, with the resonances being
formally divergent.  Such divergences are usually eliminated by
accounting for the finite lifetimes of excited Landau states.  This
practice requires the use of spin-dependent cyclotron rates in order to
obtain accurate determinations of process rates very near cyclotronic
resonances, the phase space domain most relevant for certain
applications to pulsar models.  This paper develops previous results in
the literature to obtain compact analytic expressions for cyclotron
decay rates/widths in terms of a series of Legendre functions of the
second kind; these expressions can be expediently used in astrophysical
models.  The rates are derived using two popular eigenstate formalisms,
namely that due to Sokolov and Ternov, and that due to Johnson and
Lippmann.  These constitute two sets of eigenfunctions of the Dirac
equation that diagonalize different operators, and accordingly yield
different spin-dependent cyclotron rates.  This paper illustrates the
attractive Lorentz transformation characteristics of the Sokolov and
Ternov formulation, which is another reason why it is preferable when
electron spin information must be explicitly retained.
\end{abstract}  
\section{INTRODUCTION}
 \label{sec:intro}

The number of astrophysical sources thought to possess surface magnetic
fields above the quantum critical field strength of \teq{B_{\rm
cr}=m_e^2c^3/(e\hbar ) \approx 4.413\times 10^{13}}G has been steadily
growing in recent years.  The five presently known Soft Gamma-Ray
Repeaters (SGRs) and six or seven Anomalous X-Ray Pulsars (AXPs) are
believed to be magnetars, neutron stars having surface fields in the
range $10^{14} - 10^{15}$ G (Duncan \& Thompson 1992). In addition, the
Parkes Multibeam survey (Manchester et al. 2001) has discovered at least
six new radio pulsars that have surface fields near or above critical,
and several with fields comparable to those of some of the magnetars. 
The radiation processes that determine the observed X-ray and gamma-ray
spectra in these sources are operating in the extreme relativistic and
quantum regimes and thus require treatment that is accurate in such
environments.  One process that is particularly important in the source
emission models is resonant Compton scattering, in which electrons
scatter photons at the cyclotron resonance with a cross section much
larger than at continuum energies.  Relativistic electrons can
blue-shift low energy photons into the resonance, upscattering the
photons into a high-energy continuum (Daugherty \& Harding 1989, Dermer
1990).

The quantum electrodynamical (QED) cross section for Compton scattering
at the cyclotron fundamental (Herold 1979) and higher harmonics
(Daugherty \& Harding 1986, Bussard, Alexander  \& M\'esz\'aros 1986),
accurate in arbitrarily high magnetic fields, has been known for some
time. However, these derivations do not treat the natural line widths
that render the cross section finite at the cyclotron resonances.  In
order to use such rates in spectral calculations in astrophysical
models, the line widths that originate from the lifetimes of the excited
Landau states must be included in the cross section (Wasserman \&
Salpeter 1980). The width of the $n$th resonance is equal to the
cyclotron decay rate from that state (Pavlov et al. 1991), and the
prescription for incorporating the widths in the QED cross section in
the high-field regime has been discussed by Harding \& Daugherty (1991)
and Graziani (1993).  Including resonant line widths in the scattering
cross section necessarily requires spin-dependent decay rates, which
appear in an infinite sum over Landau state $n$ and spin of the
intermediate virtual states, even in the case of ground state-to-ground
state scattering in the fundamental. Different spin states have
different decay rates, and thus different resonant energy denominators. 
Since the electrons and positrons in intermediate states have non-zero
momentum parallel to the magnetic field, it is important to use basis
states that yield a spin dependence that is Lorentz invariant, i.e.
boosts along the magnetic field do not lead to a mixing between the spin
states.

As is true for all quantum processes, the spin-dependent rates and cross
sections depend on the choice of electron wavefunctions in a uniform
magnetic field.  Historically, several choices of wavefunctions have
been used in calculations of the scattering cross section and cyclotron
decay rates.  The two most widely used wavefunctions are those of
Johnson \& Lippman (1949) and Sokolov \& Ternov (1968).  The Johnson \&
Lippman (JL) wavefunctions are derived in Cartesian coordinates and are
eigenstates of the kinetic momentum operator.  The Sokolov \& Ternov
(ST) wavefunctions, specifically their ``transverse polarization''
states, are derived in cylindrical coordinates and are eigenfunctons of
the magnetic moment operator. Given the different spin dependence of the
ST and JL eigenstates, one must use caution in making the appropriate
choice when treating spin-dependent processes.  Herold, Ruder \& Wunner
(1982) and Melrose and Parle (1983) have noted that the ST eigenstates
have desirable properties that the JL states do not possess, such as
being eigenfunctions of the Hamiltonian including radiation corrections,
having symmetry between positron and electron states, and
diagonalization of the self-energy shift operator. As found by Graziani
(1993), the ST wavefunctions also diagonalize the Landau-Dirac operator,
and are the physically correct choices for spin-dependent treatments and
for incorporating widths in the scattering cross section.  Although the
spin-averaged ST and JL cyclotron decay rates are equal, the
spin-dependent decay rates are not, except in the special case in which
the initial momentum of the electron parallel to the field vanishes.

Perhaps the most fundamental argument against use of the JL
wavefunctions is that radiative corrections cause excited JL Landau
levels to become unstable to spin-flip transition within Landau states
on timescales comparable to the timescale for decay to a lower Landau
\teq{n} state. These radiative corrections to the magnetic moment break
the spin degeneracy of both the ST and JL excited Landau states (Herold,
Ruder \& Wunner 1982), causing spin-dependent energy level shifts, and
implying a dependence of the rates for spin-flip transitions within a
Landau state on the choice of wavefunctions. The rates of such spin-flip
transitions within split ST Landau states have been evaluated (Gepr\"ags
et al. 1994; see also Parle 1987), and are found to be of order of
\teq{\fsc^6 (B/B_{\rm cr})^4} for \teq{B\ll B_{\rm cr}}, which is
negligible compared to decay rates between Landau states.  The existence
of such relatively long-lived states is a premise of the S-matrix
formalism and is also essential to any astrophysical calculations
involving spin-dependent transitions between Landau states.  In
contrast, it is anticipated that since the JL eigenstates do not
diagonalize the self-interaction Hamiltonian including radiation
corrections (i.e. the mass operator), the associated mixing incurred in
S-matrix evaluations will render spin-flip transition rates with fixed
\teq{n} comparable to ordinary cyclotronic rates with changes in
\teq{n}, a situation that is unphysical.

In this paper we discuss the Lorentz tranformation characteristics of
cyclotron decay rates for both ST and JL formulations, and derive
simplified expressions for the decay rates from an arbritrary excited
Landau state to the ground state. We show that the ST eigenstates
preserve separability of the spin dependence under Lorentz boosts along
the local magnetic field, a desirable property that does not extend to
the JL formalism, for which such Lorentz boosts mix the corresponding
spin states.  In the ST formulation, by taking advantage of its
spin-state preserving characteristics, our analytic simplifications can
be elegantly applied to the spin-dependent decay rates; in the JL
formalism the simplifications can be compactly applied only to the
spin-averaged rates. The resulting expressions replace integrals over
emergent photon angle with series of Legendre functions of the second
kind, which correspond to sums of elementary functions, easily yielding
simple asymptotic forms.

The expressions derived here should have wide applicability to modeling
cyclotron emission, Compton scattering and other QED processes in
super-critical fields.  For such magnetar-type fields, resonant
scattering takes place primarily at the fundamental, since the cyclotron
energy exceeds 1 MeV.  Although the resonant scattering line widths
formally involve infinite sums over Landau states, in the case of the
fundamental resonance the sum is dominated by the $n=1$ state, whose
width is equal to the $n \rightarrow 0$ cyclotron decay rate.  For
cyclotron scattering in higher harmonics $l > 0$, the intermediate sums
in the vertex functions have the largest contributions from the $n =
l+1$ state. However, for \teq{B\gg B_{\rm cr}}, cyclotron transitions to
the ground Landau state dominate (Sokolov, Zhukovskii \& Nikitina 1973,
White 1974, Harding \& Preece 1987), so that cyclotron decay rates for
$n \rightarrow 0$ transitions treated in this paper should be good
approximations to the widths of excited states, a circumstance pertinent
to astrophysical models of magnetars.

\section{CYCLOTRON RATES: SOKOLOV AND TERNOV FORMALISM}
 \label{sec:STformalism}
This presentation of spin-dependent cyclotron rates appropriately
focuses first on the formulation generated using the {\it transverse
polarization} eigenstates of the Dirac equation derived by Sokolov \&
Ternov (1968) for electrons in a uniform magnetic field {\bf B}. 
Explicit forms for them can also be found in Herold, Ruder \& Wunner
(1982) and Harding \& Preece (1987), whose exposition on cyclotron
emission essentially forms the basis of results developed here.  Herold
et al. observed that these states diagonalize the operator that
describes the electronic self-energy shift in an external magnetic
field.  Another attractive feature of the Sokolov \& Ternov eigenstates is
that they possess charge conjugation symmetries, i.e. between electron
(positive energy) and positron (negative energy): see Eq.~(12) of
Herold, Ruder \& Wunner (1982).  An additional asset of these wavefunctions
that is enunciated here is that they yield cyclotron rates whose spin
dependence is effectively separable from Lorentz transformations along
the field: i.e. such boosts do not mix spin states.  This is a very
useful characteristic that is not present in the Johnson \& Lippmann
formalism addressed in Section~\ref{sec:JLformalism} below.

Herold, Ruder \& Wunner (1982) obtained general expressions in their
Eq.~(17) for the spin-dependent cyclotron rate for transitions between
arbitrary Landau levels, but for the case of zero initial momentum
\teq{p} parallel to the field.  Latal (1986) independently obtained
similar results for cyclotron transitions to the ground state, but
retained arbitrary initial momenta \teq{p} along {\bf B}; his Eq.~(22)
forms the starting point for the exposition here.  Denoting \teq{\zeta
=\pm 1} as the spin quantum number of the initial electron, ground state
transitions are characterized by a single final spin \teq{\zeta'=-1}. 
The energy level quantum numbers are initially \teq{n}, and \teq{n'=0}
after the transition.  Latal's spin-dependent {\it total} rates can be
written in the form
\begin{equation}
   \Gamma^{\zeta}_{n0}\; =\; \biggl\lbrack 1
        - \dover{\zeta}{\sqrt{1+2 nB}}\; \biggr\rbrack\; \Gammave\quad ,
 \label{eq:Gamma_ST}
\end{equation}
where \teq{B} is the magnetic field strength, expressed dimensionlessly
hereafter in units of the quantum critical field \teq{B_{\rm
cr}=m_e^2c^3/(e\hbar ) \approx 4.413\times 10^{13}}Gauss, and the {\it
spin-averaged}, total cyclotron transition rate for \teq{n\to n'=0} is
given by 
\begin{eqnarray}
   \Gammave &=& \dover{\fsc c}{\lambar}\, \dover{n^n}{(n-1)!}\,
      \dover{B}{E_n} \int_{-1}^1 \dover{d\mu}{(E_n-p\mu)^2}\nonumber\\[-5.5pt]
 \label{eq:Gammave_ST}\\[-5.5pt]
      &\times & \;
      \dover{(1-\xi)^{n-1}}{(1+\xi)^{n+1}}\biggl( 1 + 2nB  - \dover{1}{\xi} \biggr)
         \; \exp \biggl( - n\, \dover{1-\xi}{1+\xi} \biggr) \; .\nonumber
\end{eqnarray}
Here \teq{\lambar = \hbar/m_ec} is the Compton wavelength of the
electron over \teq{2\pi}, and other quantities in this equation are
defined as follows. For initial and final dimensionless momenta (i.e.,
in units of \teq{m_ec}), \teq{p} and \teq{q} respectively, parallel to
the field, the electron's initial energy \teq{E_n} and final energy
\teq{E_0}, both dimensionless (i.e., in units of \teq{m_ec^2}), are
\begin{equation}
   E_n\; =\; \sqrt{1+2nB+p^2}\quad \hbox{and}\quad E_0\; =\; \sqrt{1+q^2}\quad.
 \label{eq:E_e_init_fin}
\end{equation}
This convention of using dimensionless energies and momenta will be
adopted throughout the paper.

The integration variable \teq{\mu} is the angle cosine of the emitted
photon with respect the magnetic field direction, i.e.
\teq{\mu=\cos\theta = \vec{k}.\vec{B}/\vert \vec{k}\vert\;
\vert\vec{B}\vert} for photon wavenumber vector \teq{\vec{k}}.  No
integration by parts has been performed when obtaining
Eq.~(\ref{eq:Gammave_ST}), so that the cyclotron rate, differential in
photon angles, \teq{d\Gammave /d\mu}, again averaged over initial
electron spins, corresponds directly to the integrand of
Eq.~(\ref{eq:Gammave_ST}):
\begin{equation}
   \Gammave\; =\; \int_{-1}^1 \dover{d\Gammave}{d\mu}\; d\mu\quad .
 \label{eq:Gammave_diff_def}
\end{equation}
Corresponding spin-dependent differential cyclotron rates can be deduced
from Eq.~(\ref{eq:Gamma_ST}) in similar fashion.

The remaining variable, \teq{\xi}, emerges naturally from
energy-momentum conservation in the interaction.  Only momentum parallel
to the field is conserved, since the system is not
translationally-invariant orthogonal to {\bf B}.  Hence cyclotron
emission satisfies two such conservation relations:
\begin{eqnarray}
   E_n &=& E_0 + \omega\quad ,\nonumber\\[-5.5pt]
 \label{eq:E_p_conserve}\\[-5.5pt]
   p &=& q + \omega\mu\nonumber
\end{eqnarray}
for \teq{\mu =\cos\theta}.  Here \teq{\omega} is the dimensionless 
photon energy, having been scaled by \teq{m_ec^2}.  The 
elimination of \teq{q} from the system in Eq.~(\ref{eq:E_p_conserve}) 
leads to the identities
\begin{eqnarray}
   \omega &=& \dover{2nB}{1+\xi}\, \dover{1}{E_n - p\mu}\quad ,\nonumber\\[-5.5pt]
\label{eq:omega_ident}\\[-5.5pt]
   \omega (1-\mu^2) &=& (1-\xi )\, (E_n - p\mu )\nonumber
\end{eqnarray}
that are simultaneously satisfied, with
\begin{equation}
   \xi \; =\; \sqrt{1 - \dover{2nB\, (1-\mu^2)}{(E_n - p\mu )^2}} \quad ,
 \label{eq:xi_def}
\end{equation}
being the \teq{n'\to 0} specialization of the definition in Eq.~(11b) of 
Latal (1986).  The angle integration in Eq.~(\ref{eq:Gammave_ST}) for the 
cyclotron rate is non-trivial, due largely to the complicated dependence of 
\teq{\xi} on \teq{\mu}, and the presence of the exponential.

This completes the definitions relevant to Eq.~(\ref{eq:Gammave_ST}).
However, we also note that the identities in Eq.~(\ref{eq:omega_ident})
combine to yield
\begin{equation}
   \kappa\;\equiv\; \dover{\omega^2 (1-\mu^2)}{2B}
               \; =\; n\, \dover{1-\xi}{1+\xi}\quad ,
 \label{eq:kappa_def}
\end{equation}
which is precisely the argument of the exponential in
Eq.~(\ref{eq:Gammave_ST}).  This \teq{\exp (-\kappa )} factor results
from the \teq{n'=0} specialization of associated Laguerre functions that
are formed from the spatial integration (Fourier transform) of the
eigenfunctions. Since \teq{\omega\sin\theta} is Lorentz invariant under
boosts along {\bf B}, so also is \teq{\kappa}.  Furthermore, forming
\teq{E_n-p\mu} using Eq.~(\ref{eq:E_p_conserve}), and then eliminating
\teq{\omega} using the second identity in Eq.~(\ref{eq:omega_ident}), 
quickly leads to the equivalence
\begin{equation}
   \xi\; =\; \dover{E_0-q\mu}{E_n-p\mu}\quad .
 \label{eq:xi_altern}
\end{equation}
When weighted by \teq{\omega} top and bottom, this is a ratio
of invariant products of four-momenta associated with the interaction; 
i.e. \teq{\xi} is also a Lorentz invariant for boosts along the field, which
can be inferred directly from Eq.~(\ref{eq:kappa_def}).

\subsection{Lorentz Transformation Properties}
 \label{sec:ST_Lorentz}
There is no need to compute Eq.~(\ref{eq:Gammave_ST}) as it stands for
arbitrary \teq{p}.  Instead, one can appeal to a simple Lorentz
transformation protocol.  Consider boosts along the field between the
inertial frame where the initial electron possesses momentum \teq{p}
along the field, and that frame where the initial parallel momentum is
zero. In this latter ``rest'' frame let \teq{\mu_0} be the photon angle
cosine with respect to {\bf B}, \teq{p_0=0} be the initial parallel
\teq{e^-} momentum, and the initial electron energy be denoted
\begin{equation}
   \erg_n\; =\; \sqrt{1+2 nB}\quad .
 \label{eq:erg_n_def}
\end{equation}
Then, for a dimensionless boost velocity \teq{\beta \equiv v/c} and Lorentz
factor \teq{\gamma =1/\sqrt{1-\beta^2}} along {\bf B}, the initial electron 
quantities in Eq.~(\ref{eq:Gammave_ST}) satisfy
\begin{eqnarray}
   E_n &=& \gamma (\erg_n -\beta p_0)
            \;\equiv\; \gamma\erg_n\quad , \nonumber\\[-5.5pt]
 \label{eq:Lorentz_boost}\\[-5.5pt]
   p &=& \gamma (p_0 - \beta \erg_n)
            \;\equiv\; -\gamma\beta\erg_n\quad , \nonumber
\end{eqnarray}
from which  \teq{\beta = -p/E_n =-p/\sqrt{1+2nB+p^2}} is established.
The photon angle cosine satisfies the aberration formula
\begin{equation}
   \mu\;\equiv\;\cos\theta\; =\; \dover{\mu_0 -\beta}{1-\beta\mu_0}\quad .
 \label{eq:aberration}
\end{equation}
This latter relation identifies a suitable change of variables for the
angle integration, for which \teq{d\mu =d\mu_0/[\gamma
(1-\beta\mu_0)]^2}. Other transformation identities include
\teq{\omega\sin\theta = \omega_0\sin\theta_0} and \teq{(1-\mu^2)/(E_n
-p\mu )^2 = (1-\mu_0^2)/\erg_n^2}, from which the invariance in form of
\teq{\xi} is established via the definition in Eq.~(\ref{eq:xi_def}).  
It then follows, after a modicum of algebra, that this pure change of 
variables leads to an alternative form for the rate in Eq.~(\ref{eq:Gammave_ST}):
\begin{equation}
   \Gammave \; =\; \dover{\fsc c}{\lambar}\, \dover{n^n}{(n-1)!}\,
      \dover{B}{\gamma\,\erg_n} \; I_n(B)\quad ,
 \label{eq:Gammave_ST_alt}
\end{equation}
with
\begin{equation}
      I_n(B)\; =  \int_{-1}^1 \dover{d\mu_0}{\erg_n^2}
      \dover{(1-\xi)^{n-1}}{(1+\xi)^{n+1}}\Bigl( \erg_n^2  - \dover{1}{\xi} \Bigr)
         \,\exp \biggl( - n\, \dover{1-\xi}{1+\xi} \biggr) \; ,
 \label{eq:InB_def}
\end{equation}
where, now \teq{\xi} is obtained from Eq.~(\ref{eq:xi_def}) using the
correspondences \teq{\mu\to\mu_0}, \teq{p\to 0} and \teq{E_n\to\erg_n}.

Accordingly, this identifies attractive Lorentz transformation behavior,
with the rate reduced (i.e. lifetime dilated) by just the Lorentz factor
\teq{\gamma} of the boost.  Moreover, such boosts along the field keep
the spin states ``separated,'' i.e. there is no implied mixing of states
incurred by such transformations. This inherent simplicity is an
appealing characteristic of the Sokolov \& Ternov states, and was
identified by Graziani (1993); it is not exhibited in the Johnson \&
Lippman formalism explored in Sec.~\ref{sec:JLformalism}.  Clearly, this
extraction of \teq{p >0} cases via a simple modification factor outside
the integral is expedient for the subsequent analytic developments.

\subsection{Analytic Developments}
 \label{sec:ST_analytic}

While the integral expressions for the cyclotron rates can be routinely
evaluated numerically, they can also be represented by compact analytic 
series in terms of elementary functions that are readily amenable to 
computation.  The integration variable \teq{\mu_0} for \teq{I_n(B)} in
Eq~(\ref{eq:InB_def}) is not the most convenient; a more expedient
choice for the purposes of analytic reduction is
\begin{equation}
   \phi\;\equiv\;\dover{\kappa}{n}\; =\;\dover{1-\xi}{1+\xi}\quad ,\quad
   \xi\; =\; \sqrt{\dover{1+2nB\,\mu_0^2}{1+2nB}}\quad .
 \label{eq:phi_def}
\end{equation}
The integration is even in \teq{\mu_0}, and since \teq{1/\erg_n\leq
\xi\leq 1}, the integration range maps over to
\begin{equation}
   0\;\leq\; \phi\;\leq\; \phi_n\;\equiv\; \dover{\erg_n -1}{\erg_n +1}\quad .
 \label{eq:phin_def}
\end{equation}
After a modest amount of algebra, the change of variables then leads to
the form
\begin{equation}
   I_n(B)\; = \int_0^{\phi_n} \dover{d\phi \, e^{-n\phi}\, \phi^{n-1}}{
        \sqrt{(\phi_n -\phi )\, (1/\phi_n - \phi )}}\;
        \Biggl\lbrack 1 - \dover{\phi}{2} \biggl( \phi_n +
                \dover{1}{\phi_n}\, \biggr)\, \Biggr\rbrack\quad .
 \label{eq:InB_alt}
\end{equation}
This integral can be evaluated in terms of Appell functions, the
degenerate, two-dimensional hypergeometric functions, using Eq.~3.385 of
Gradshteyn \& Ryzhik (1980).  Such a step does not facilitate
evaluation, since Appell function expositions usually develop double,
infinite power series representations (e.g. see Burchnall \& Chaundy 1941;
Exton 1976).

Analytic progress is quickly made via Taylor series expansion of the
exponential \teq{e^{-n\phi}} around \teq{\phi=0}.  The order of integration
and the infinite summation can then be interchanged because the integration 
is uniformly convergent on the interval \teq{0\leq\phi\leq \phi_n}, since 
\teq{\phi_n} is strictly less than unity.  Another change of variables 
\teq{\phi =e^{-t}}, with the definition
\begin{equation}
   z_n\; =\; \cosh t_n
          \; =\; \dover{1}{2} \biggl( \phi_n + \dover{1}{\phi_n} \biggr)
          \;\equiv\; 1 + \dover{1}{nB}\quad ,
 \label{eq:zn_def}
\end{equation}
then establishes
\begin{equation}
   I_n(B)\; =\; \sum_{k=0}^{\infty} \dover{(-n)^k}{k!}\, J_{n+k}(z_n)\quad ,
 \label{eq:InB_alt_too}
\end{equation}
with
\begin{equation}
   J_{\nu}(z_n)\; =\; \dover{1}{\sqrt{2}} \int_{t_n}^{\infty}
      \dover{dt\, e^{-(\nu -1/2)t}}{\sqrt{\cosh t - \cosh t_n}}\;
      \Bigl\lbrack 1 - e^{-t}\cosh t_n \Bigr\rbrack\;\; .
 \label{eq:Jzn_def}
\end{equation}
Using the identity 8.715.2 of Gradshteyn \& Ryzhik
(1980) gives an integral representation of the Legendre
function \teq{Q_{\nu}(z_n)} of the second kind:
\begin{equation}
   Q_{\nu}(z_n)\; =\; \dover{1}{\sqrt{2}} \int_{t_n}^{\infty}
      \dover{dt\, e^{-(\nu +1/2)t}}{\sqrt{\cosh t - \cosh t_n}}
      \quad ,\quad z_n\; >\; 1\quad .
 \label{eq:Qnu_def}
\end{equation}
These special functions are finite sums of elementary logarithmic and
polynomial functions of \teq{z_n} (e.g. see Abramowitz \& Stegun 1965).
Note that the \teq{Q_{\nu}(z)} are generalizable to associated Legendre
functions \teq{Q_{\nu}^{\mu}(z)}, defined in 8.703 of Gradshteyn \&
Ryzhik (1980), from which \teq{Q_{\nu}(z)\equiv Q_{\nu}^0(z)}. It
follows from Eqs.~(\ref{eq:Jzn_def}) and~(\ref{eq:Qnu_def}) that
\begin{equation}
   J_{n+k}(z_n)\; =\; Q_{n+k-1}(z_n) - z_n Q_{n+k}(z_n)\quad ,
 \label{eq:Jnu_ident}
\end{equation}
and relevant properties of the \teq{Q_{\nu}} and \teq{J_{\nu}} functions
are listed in the Appendix.  Computational issues for the \teq{J_m} are 
discussed there also, and an expedient approach for integer \teq{m=n+k}
is to compute the \teq{J_m} using the recurrence relation in
Eq.~(\ref{eq:Jrecurrence}).  The final result of these analytic reductions is 
\begin{equation}
   \Gammave \; =\; \dover{\fsc c}{\lambar}\, \dover{n^n}{(n-1)!}\,
      \dover{B}{\gamma\erg_n} \; 
      \sum_{k=0}^{\infty} \dover{(-n)^k}{k!}\, J_{n+k}(z_n) \quad ,
 \label{eq:Gammave_ST_red}
\end{equation}
which expresses the spin-averaged cyclotron rate as a comparatively 
simple series of elementary functions.  For low values of \teq{n}, this
series is rapidly convergent for all \teq{z_n\geq 1}, typically requiring
3--4 terms to achieve 0.01\% accuracy.  Such a rate of convergence
renders the series evaluation computationally much more efficient than 
a numerical integration of Eq.~(\ref{eq:Gammave_ST}) by Simpson's rule 
or a quadrature technique.  The appearance of the Legendre 
functions of the second kind,  \teq{Q_{\nu}(z)}, in the series expansion 
for the rate is a result of the cylindrical symmetry imposed by the 
presence of the external magnetic field.

\figureoutpdf{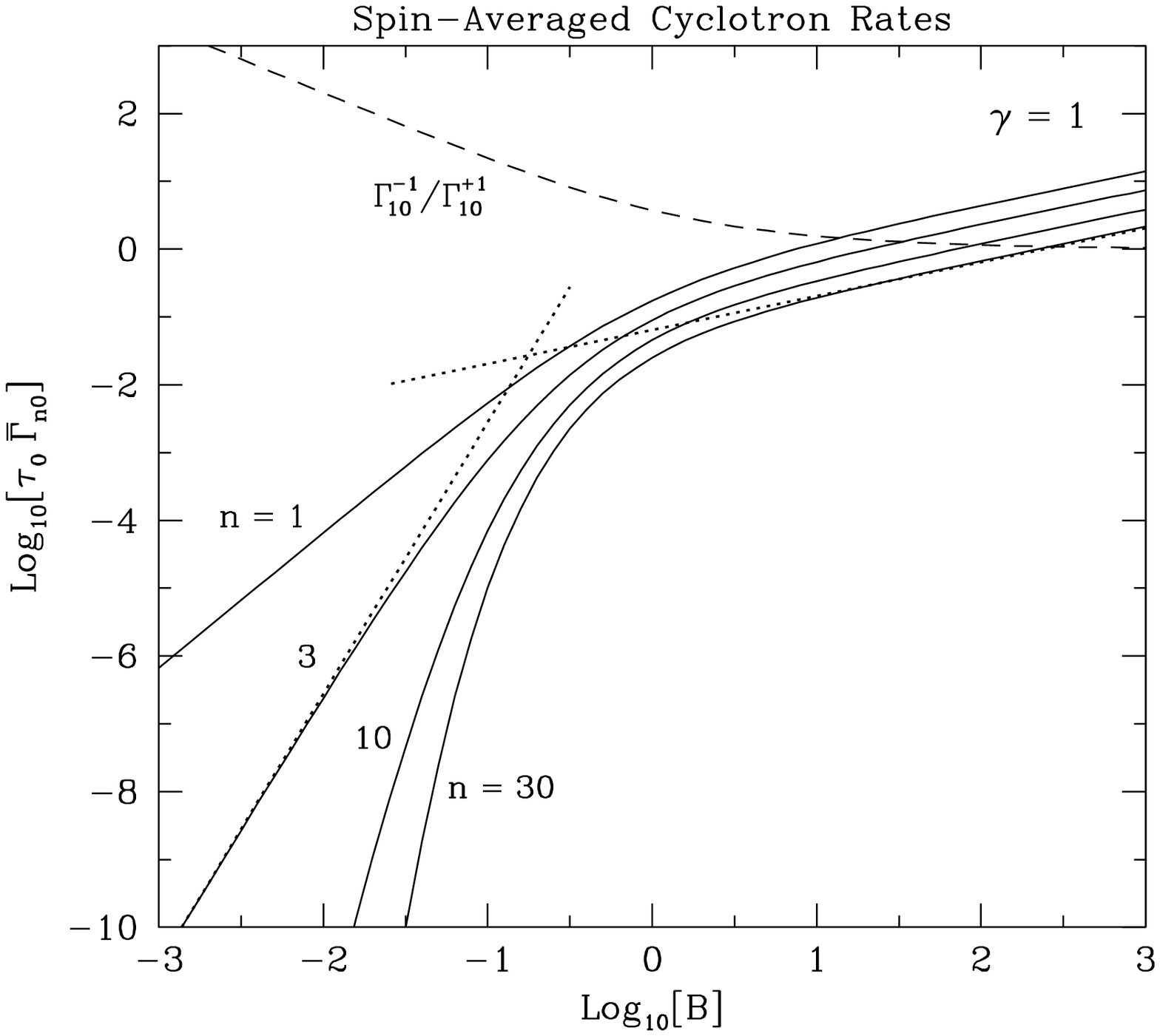}{4.0}{0.0}{-0.2}{
The spin-averaged cyclotron decay rates \teq{\Gammave} (solid curves)
resulting from Eq.~(\ref{eq:Gammave_ST_red}), or equivalently
Eq.~(\ref{eq:Gammave_ST}), for different values of \teq{n}, as labelled.
The rates are scaled by the typical lifetime \teq{\tau_0=\fsc
c/\lambar}, and are plotted as functions of the magnetic field strength
\teq{B}, expressed in units of \teq{B_{\rm cr}=4.413\times
10^{13}}Gauss.  The dotted lines represent the power-law asymptotic
dependences (i) in the limit \teq{n B\ll 1} in
Eq.~(\ref{eq:Gammave_lownB}), for \teq{n=3}, and (ii) in the \teq{n B\gg
1} limit in Eq.~(\ref{eq:Gammave_highnB_too}), specifically for
\teq{n=30}.  The dashed curve is the logarithm of the ratio of 
spin-dependent rates for \teq{n=1} in the Sokolov \& Ternov formalism, i.e., 
Eq.~(\ref{eq:Gamma_ratio}).  Results are presented for zero electron
momentum initially along {\bf B}, i.e., \teq{p=0} or  \teq{\gamma=1}.
 \label{fig:cyc_ST_ave}} 

Evaluations of Eq.~(\ref{eq:Gammave_ST_red}) are presented in
Fig.~\ref{fig:cyc_ST_ave} for different values of \teq{n} and for
\teq{\gamma =1}, demonstrating the well-known rapid increase of the rate
with \teq{B} when \teq{B\lesssim 1}.  The dependence on \teq{B} is
relatively weak for supercritical fields.  The Figure also clearly
exhibits the property \teq{d\Gammave /dn < 0}, which can, with extensive
effort, be deduced from the integral form for the spin-averaged rate in
Eq.~(\ref{eq:Gammave_ST_alt}). This decline of \teq{\Gammave} with
increasing \teq{n} is evident in the asymptotic approximations for
\teq{n B\ll 1} and \teq{n B\gg 1} derived shortly.
Fig.~\ref{fig:cyc_ST_ave} also exhibits the logarithm of the ratio
\begin{equation}
   \dover{\Gamma^{-1}_{n0}}{\Gamma^{+1}_{n0}}\; =\;
   \dover{\sqrt{1+2 n B}+1}{\sqrt{1+2 n B}-1}
 \label{eq:Gamma_ratio}
\end{equation}
of the spin-dependent rates for \teq{n=1}; higher \teq{n} cases display
similar behavior.  This curve illustrates how the rate is approximately
independent of initial spin when \teq{n B\gg 1}, whereas \teq{\zeta =-1}
initial states lead to much faster transition rates (by of order
\teq{(nB)^{-1}}) when \teq{n B\ll 1}, reflecting the well-known
dominance of non-spin-flip cyclotron transitions by non-relativistic
electrons (e.g. see Melrose \& Zheleznyakov 1981).

\subsubsection{Asymptotic Limits}
 \label{sec:STasymptotics}

Now consider the two pertinent asymptotic limits for the spin-averaged
rate in Eq.~(\ref{eq:Gammave_ST_red}). When \teq{n B\ll 1},
\teq{\phi_n\approx nB/2\ll 1} and \teq{z_n\gg 1}. Using the first
identity in Eq.~(\ref{eq:J_nu_ident}), Eq.~8.771.2 of Gradshteyn \&
Ryzhik (1980) can then be invoked to obtain the leading order dependence
of \teq{J_n(z_n)}, noting the specialization \teq{{}_2F_1(\alpha ,\,
\beta ,\, \gamma ,\, 0)=1} of the common hypergeometric function.  The
result, for general arguments \teq{z}, is
\begin{equation}
   J_n(z)\;\approx\; \dover{ (n+1)\,\Gamma (n)\, \Gamma(1/2)}{
     2\, \Gamma (n+3/2)}\; (2 z)^{-n}\quad ,\quad z\;\gg\; 1\quad .
 \label{eq:J_n_lownB}
\end{equation}
Here, \teq{\Gamma (x)} represents the Gamma function.  This is the 
leading order (\teq{k=0}) term in the series evaluation of \teq{I_n(B)}, 
and hence this result is readily checked by taking the \teq{\phi_n\to 0}
limit of Eq.~(\ref{eq:InB_alt}).  The doubling formula for the Gamma
function, in Eq.~8.335.1 of Gradshteyn \& Ryzhik (1980), can now be
used to yield an asymptotic form
\begin{equation}
   \Gammave \; \approx\; \dover{\fsc c}{\lambar}\, (2n^2)^n\,
   \dover{(n+1)!}{(2n+1)!}\, B^{n+1}\quad ,\quad nB\;\ll\; 1\quad ,
 \label{eq:Gammave_lownB}
\end{equation}
which is just Eq.~(31d) of Latal (1986).  This form illustrates the
strong dependence of the cyclotron rates on both \teq{n} and \teq{B}. A
sample \teq{n=3} case of the limit in Eq.~(\ref{eq:Gammave_lownB}) is
displayed in Fig.~\ref{fig:cyc_ST_ave}. This limit can only be compared
with the well-studied classical cyclotron limit in a restricted fashion,
since it is still an essentially quantum result. Moreover, it
incorporates the intrinsically relativistic effect of spin-orbit
coupling.  Observe that Eq.~(\ref{eq:Gammave_lownB}), in conjunction
with Eq.~(\ref{eq:Gamma_ST}), reproduces both the non-spin-flip
(\teq{\zeta =-1}) and spin-flip (\teq{\zeta =1}) ST rates encapsulated
in Eqs.~(18) and~(21), respectively, of Melrose \& Russell (2002), or
equivalently, the \teq{n'=0} specialization of Eq.~(18) of Herold, Ruder
\& Wunner (1982).

Classical formulations of cyclotron emissivities (e.g. see Eq.~(6.19) of
Bekefi 1966) yield rates that depend on components \teq{\beta_{\perp}c}
of the electron velocity perpendicular to the field:
\teq{\Gamma_n\propto B\beta_{\perp}^{2n}}.  Here the classical rate

\teq{\Gamma_n} effectively represents the \teq{nB\ll 1} limit of a sum
over various \teq{n\to n'} transitions, with \teq{n'<n}.  By invoking
the correspondence \teq{2nB\to p_{\perp}^2 =\beta_{\perp}^2}, it is
quickly established that \teq{\Gamma_n\propto B^{n+1}}, the same
dependence as in Eq.~(\ref{eq:Gammave_lownB}).  Yet in general, the
classical \teq{\Gamma_n} does not equal the quantum \teq{\Gammave},
since here \teq{n'>0} contributions are omitted from consideration.
However, in the particular case of \teq{n=1}, there is only a single
\teq{n'<n} final state, namely \teq{n'=0}, and with the prescription
\teq{\beta_{\perp}^2 \to 2B}, Eq.~(6.19) of Bekefi (1966) specializes to
yield \teq{\Gamma_1=4\fsc c\, B^2/(3\lambar )}, which is precisely the
value obtained in Eq.~(\ref{eq:Gammave_lownB}). Derivation of the \teq{n
B\gg 1}, ultra-quantum limit is slightly more involved. In this
situation, \teq{\phi_n\to 1} and \teq{z_n\to 1} so that the entire
\teq{k} series in Eq.~(\ref{eq:Gammave_ST_red}) is retained. The key
useful result is derived in the Appendix, in
Eq.~(\ref{eq:J_n_Taylor_zeq1}), which establishes \teq{J_{n+k}(z)\to
1/(n+k)} as \teq{z\to 1^+}.  Including the next order logarithmic
contribution then leads to an approximate form for \teq{I_n(B)} in
Eq.~(\ref{eq:InB_Taylor_series}), so that it then follows that 
\begin{equation}
   \Gammave \;\approx \; \dover{\fsc c}{\lambar} \, \sqrt{\dover{B}{2n}} 
       \, \Biggl\{ \dover{\gamma (n,\, n)}{\Gamma (n)}
          - \dover{n^ne^{-n}}{\Gamma (n)}\, \dover{\log_e2nB}{2nB}\, \Biggr\}
       \; ,\;\; nB\,\gg\, 1\; .
 \label{eq:Gammave_highnB}
\end{equation}
Here, \teq{\gamma (n,\, x)} is the incomplete Gamma function in 8.350.1
of Gradshteyn \& Ryzhik (1980).  Note that the same result can be
derived directly from the \teq{\phi_n\to 1} limit of
Eq.~(\ref{eq:InB_alt}) using the integral definition of the incomplete
Gamma function. The weak \teq{\sqrt{B}} dependence of the rate on
\teq{B} in this limit is now apparent, and is evinced in
Fig.~\ref{fig:cyc_ST_ave} by the crowding of curves for supercritical
fields. In the limit of \teq{n\gg 1}, further specialization is possible
using the limit \teq{\gamma (n, \, n)/\Gamma(n)\to 1/2} as
\teq{n\to\infty}, derived in the Appendix, to yield the approximation
\begin{equation}
   \Gammave\;\approx\; \dover{\fsc c}{\lambar}\, \sqrt{\dover{B}{8n}}
   \quad ,\quad n\;\gg\; 1\;\; ,\;\; B\;\gg\; 1\quad .
 \label{eq:Gammave_highnB_too}
\end{equation}
This ultra-quantum result corresponds to the limit given in Eq.~(32a) of
Latal (1986), though the numerical coefficient of Latal's does not agree
exactly with the \teq{1/\sqrt{8}} factor in
Eq.~(\ref{eq:Gammave_highnB_too}), the difference being around 10\%. 
However, the expressions for \teq{n\to 0} quantum cyclotron transition
rates derived in Eq.~(9) of Sokolov, Zhukovskii \& Nikitina (1973) and
Eq.~(2.56) of White (1974) reduce exactly to
Eq.~(\ref{eq:Gammave_highnB_too}) in the limit \teq{B\gg 1}, confirming
the asymptotic analysis here.  A sample \teq{n=30} case of this limit is
displayed in Fig.~\ref{fig:cyc_ST_ave}, the precision of which is around
2\% relative to the exact result, and better than Eq.~(32a) of Latal
(1986).  The first order correction to this asymptotic limit
that would follow from the logarithmic term in
Eq.~(\ref{eq:Gammave_highnB}) is \teq{O(\log_en/\sqrt{n})} relative to
the leading order contribution in Eq.~(\ref{eq:Gammave_highnB_too}).

\subsubsection{Differential Photon Spectra}
 \label{sec:diff_spectrum}

In the \teq{p=0} frame, the spin-averaged differential cyclotron rates
\teq{d\Gammave /d\mu_0} can be immediately extracted from
Eq.~(\ref{eq:Gammave_ST_alt}) using the prescription implied by
Eq.~(\ref{eq:Gammave_diff_def}).  However, given that
Eqs.~(\ref{eq:omega_ident}) and~(\ref{eq:xi_def}) define a one-to-one
correspondence between the angle cosine \teq{\mu_0} and the energy
\teq{\omega} of the emitted cyclotron photon, the differential rates can
suitably be presented as differential photon spectra, \teq{d\Gammave
/d\omega}.  The photon energy is \teq{\omega = (\erg_n-1/\erg_n)/(1+\xi
)}, so that the kinematic bound \teq{1/\erg_n\leq \xi\leq 1} translates
to 
\begin{equation}
   \dover{nB}{\sqrt{1+2nB}} \;\equiv\; \omega_- 
   \;\leq\;\omega\; \;\leq\; \omega_+\;\equiv\;\sqrt{1+2nB} -1\quad .
 \label{eq:omega_bounds}
\end{equation}
The change of variables is then defined by two branches,
\begin{equation}
   \mu_0\; =\; \pm \dover{1}{\omega}\, \sqrt{(\erg_n-1-\omega )(\erg_n+1-\omega )}\quad ,
 \label{eq:mu0_omega_rel}
\end{equation}
with the factors inside the square root reflecting the kinematics
associated with both the cyclotron radiation and the one-photon pair
creation interactions.  This has a Jacobian \teq{d\mu_0/d\omega = -\xi
(1+\xi) \erg_n^2 /[(\erg_n^2-1)\mu_0\omega ]}, leading to the
spin-averaged differential spectrum
\begin{eqnarray}
   \dover{d\Gammave}{d\omega} &=& \dover{\fsc c}{\lambar}\, 
         \dover{1}{B\, \erg_n^2}\, \dover{n^{n-1}\,  e^n}{\Gamma (n+1)}
         \; \Bigl( \dover{\omega}{\omega_-} -1\Bigr)^{n-1} \nonumber\\[-5.5pt]
 \label{eq:dGammave_domega_ST}\\[-5.5pt]
      &\times & 
         \dover{(\erg_n-\omega )nB-\omega}{
            \sqrt{(\erg_n-1-\omega )(\erg_n+1-\omega )}}
         \; \exp \Bigl(  - n\, \dover{\omega}{\omega_-} \Bigr) \; .\nonumber
\end{eqnarray}
This spectral rate is summed over the polarizations of the radiated
photons. Note that a factor of \teq{2} has been included to account for
the two branches of the relation between \teq{\mu_0} and \teq{\omega}. 
Observe also that this spectrum can be obtained directly from
Eq.~(\ref{eq:Gammave_ST_alt}) and the form for \teq{I_n(B)} in
Eq.~(\ref{eq:InB_alt}) by using the correspondence \teq{\phi =
\omega/\omega_- -1}.

Key characteristics of the differential spectrum include its broadening
and shifting to higher energies as \teq{nB} increases, behavior that is
largely governed by the kinematic result in Eq.~(\ref{eq:omega_bounds}).
 An exponential decline is present in \teq{n=1} cases until a sharp
kinematic peaking at \teq{\omega\sim\omega_+=\erg_n-1} overtakes it.  In
\teq{n \gg 1} cases, this exponential contribution is dominated by the
power-law factor \teq{(\omega /\omega_--1)^{n-1}}, leading to a
steeply-rising spectrum. When \teq{nB\gg 1}, nearly all of the radiative
power in the spectrum is confined to energies \teq{\omega\approx
\sqrt{2nB}}. In the opposite domain, \teq{nB\ll 1}, the spectra are
extremely narrow, and little insight is gained by attempting an involved
comparison with the classical limit; most classical results are rendered
for sums over a multitude of transitions \teq{n\to n'}.

\subsubsection{Cyclotron Power}
 \label{sec:power_ST}

To conclude the formalism based on Sokolov \& Ternov
states, expressions for the cyclotron energy loss rate, or power,
\teq{P_{n0}=dE_{n0}/dt} can be expeditiously presented.  The development
of such rates parallels that for \teq{\Gammave}, with the inclusion of
an extra (dimensionless) weighting factor \teq{\omega} to the
differential decay rate \teq{d\Gammave /d\mu}.  Hence, one can quickly
write down
\begin{eqnarray}
   \Powerave &\equiv& m_ec^2 
   \int_{-1}^1\omega\, \dover{d\Gammave}{d\mu}\, d\mu\nonumber\\[2pt]
    &=& \dover{\fsc m_ec^3}{\lambar}\, \dover{n^n}{(n-1)!}\,
      \dover{B}{E_n} \int_{-1}^1 \dover{\omega \, d\mu}{(E_n-p\mu)^2}
 \label{eq:Powerave_ST}\\[2pt]
      &\times & \;
      \dover{(1-\xi)^{n-1}}{(1+\xi)^{n+1}}\biggl( 1 + 2nB  - \dover{1}{\xi} \biggr)
         \; \exp \biggl( - n\, \dover{1-\xi}{1+\xi} \biggr) \; .\nonumber
\end{eqnarray}
as the spin-averaged power, using Eq.~(\ref{eq:Gammave_ST}).  This can
be conveniently transformed to the integration variable \teq{\mu_0},
representing the \teq{p=0} frame, using an analysis similar to that
presented in Sec.~\ref{sec:ST_Lorentz}.  The only change is the
manipulation of the additional \teq{\omega =2nB/(1+\xi )/(E_n-p\mu )}
factor inside the integration, for which the Lorentz transformation relations 
in Eqs.~(\ref{eq:Lorentz_boost}) and~(\ref{eq:aberration}) quickly yield
\teq{1/(E_n-p\mu )=\gamma (1-\beta\mu_0)/ \erg_n}.  Since 
\teq{d\Gammave/d\mu_0} is even in \teq{\mu_0}, it follows that the term 
proportional to \teq{\mu_0\,d\Gammave/d\mu_0} in the transformed integrand 
of Eq.~(\ref{eq:Powerave_ST}) does not contribute.  Therefore, comparison
with Eqs.~(\ref{eq:Gammave_ST_alt}) and~(\ref{eq:InB_def}) then
indicates that
\begin{equation}
   \Powerave \; =\; \dover{\fsc m_ec^3}{\lambar}\, \dover{n^{n+1}}{(n-1)!}\,
      \dover{B^2}{\erg_n^2} \; K_n(B)\quad ,
 \label{eq:Powerave_ST_alt}
\end{equation}
is the form for the cyclotron power for general \teq{p\geq 0}, with
\begin{equation}
      K_n(B)\; =  2 \int_{-1}^1 \dover{d\mu_0}{\erg_n^2}
      \dover{(1-\xi)^{n-1}}{(1+\xi)^{n+2}}\Bigl( \erg_n^2  - \dover{1}{\xi} \Bigr)
         \,\exp \biggl( - n\, \dover{1-\xi}{1+\xi} \biggr) \; ,
 \label{eq:KnB_def}
\end{equation}
where \teq{\xi} is given by Eq.~(\ref{eq:phi_def}).  This expression for
the power is identical to that in Eq.~(28) of Latal (1986), who
illustrated the magnetic field and \teq{n} dependence in his Fig.~1. 
The spin-dependent powers \teq{P_{n0}^{\zeta}} are, of course, just
given by additional multiplicative factors: \teq{P_{n0}^{\zeta} =
(1-\zeta/\erg_n)\Powerave} for \teq{\zeta =\pm 1}.

Observe that the \teq{\gamma} dependence has now disappeared from the
analysis, highlighting the fact that \teq{\Powerave} is a Lorentz invariant for
boosts along {\bf B}.  This follows from the exact compensation of time
dilation and photon redshifting effects when integrating over all
angles; such an attractive invariance property for the total power
clearly does not extend to its differential counterpart
\teq{d\Powerave/d\mu}.

The integral for \teq{K_n(B)} can be manipulated in similar fashion to
that for \teq{I_n(B)}.  Since the only difference is an extra
\teq{2/(1+\xi )} factor, which translates to a \teq{(1+\phi )} factor
in the light of Eq.~(\ref{eq:phi_def}), it follows that \teq{K_n(B)} can
be worked into the form of Eq.~(\ref{eq:InB_alt}), but with the extra
\teq{(1+\phi )} factor.  The subsequent developments in
Sec.~\ref{sec:ST_analytic} rapidly lead to the identity
\begin{eqnarray}
   K_n(B) &=& \sum_{k=0}^{\infty} \dover{(-n)^k}{k!}\, 
             \Bigl\{ J_{n+k}(z_n) + J_{n+k+1}(z_n) \Bigr\} \nonumber\\[-5.5pt]
 \label{eq:KnB_alt_too}\\[-5.5pt]
            &\equiv & \sum_{k=0}^{\infty} \dover{(-n)^k}{k!}\, 
             \biggl( 1 - \dover{k}{n} \biggr) J_{n+k}(z_n) \;\; ,\nonumber
\end{eqnarray}
for insertion into Eq.~(\ref{eq:Powerave_ST_alt}).  Given this form,
asymptotic limits can be routinely obtained by replicating the
approaches of Sec.~\ref{sec:STasymptotics}.  When \teq{nB\ll 1}, it is
evident that \teq{K_n(B)\approx I_n(B)}, since then \teq{J_{n+1}(z_n)\ll
J_n(z_n)}, as can be inferred for \teq{z_n\gg 1} from
Eq.~(\ref{eq:J_n_lownB}).  Furthermore, for \teq{nB\gg 1}, the result in
Eq.~(\ref{eq:J_n_Taylor_zeq1}) quickly yields a leading order
contribution \teq{K_n(B)\approx \gamma (n, \, n)/n^n + \gamma (n+1, \,
n)/n^{n+1}}.  Then the identity \teq{\gamma (n+1,\, n)=n\, \gamma(n, \,
n) - n^n e^{-n}} and Eq.~(\ref{eq:gamma_inc_limit}) can be used to
derive the limiting form as \teq{n\to\infty}.  The results are
\begin{equation}
   \Powerave \; \approx\; \dover{\fsc m_ec^3}{\lambar}
       \cases{2^n\, n^{2n+1}\, \dover{(n+1)!}{(2n+1)!}\, B^{n+2}\; , &
            $\!\! nB\;\ll\; 1\;$,\cr
           \dover{B}{2}\; ,&  {\hskip -10pt} $\!\! n,\; B\;\gg\; 1\;$.}
 \label{eq:Powerave_asymptotics}
\end{equation}
While the \teq{nB\ll 1} case here replicates that inferred from Eq.~(31)
of Latal (1986), the \teq{nB\gg 1} result differs from the numerical
result in Eq.~(32a) of Latal (1986) by around 10\%.

The mean cyclotron photon energy \teq{\langle\omega\rangle} is just the
ratio of \teq{\Powerave} to \teq{\Gammave}, so that one quickly deduces
from a comparison of Eq.~(\ref{eq:Powerave_asymptotics}) with
Eqs.~(\ref{eq:Gammave_lownB}) and~(\ref{eq:Gammave_highnB_too}) that
\teq{\langle\omega\rangle \approx nB} when \teq{nB\ll 1}, as expected
from classical cyclotron theory, and that \teq{\langle\omega\rangle
\approx \sqrt{2nB}} when \teq{nB\gg 1}.  Note that Latal (1986) misses
the \teq{\sqrt{2}} factor (see his Eq.~(32b) for the \teq{nB\gg 1}
domain) due to the imprecise nature of his estimates of the pertinent
numerical factors in \teq{\Powerave} and \teq{\Gammave}. These behaviors
of \teq{\langle\omega\rangle} are reflected by the differential spectral
form in Eq.~(\ref{eq:dGammave_domega_ST}), where for \teq{nB\ll 1} the
spectrum is essentially a delta function pinned at the cyclotron
harmonic energy \teq{nB}, and where for \teq{n\gg 1} and  \teq{B\gg 1}
the spectrum is somewhat broad but markedly asymmetric, skewed strongly
towards \teq{\omega\approx\omega_+\approx \sqrt{2nB}}.

\section{THE JOHNSON AND LIPPMANN FORMULATION}
 \label{sec:JLformalism}

A complementary formulation of the cyclotron problem was provided by
Daugherty \& Ventura (1978, hereafter DV78) using wavefunctions derived
by Johnson \& Lippmann (1949).  This is essentially the foremost
presentation in the literature of cyclotron radiation calculations that
specifically uses the Johnson \& Lippmann eigenstates, hereafter
referred to as JL states.  These states are eigenvalues simultaneously
of the Dirac Hamiltonian and the kinetic momentum operator
\teq{\vec{\pi} = \vec{p} + e \vec{A}/c}, where \teq{\vec{A}} is the
vector potential, and therefore are not eigenstates of any physically
meaningful spin operator (see Melrose \& Parle 1983). The relative lack
of symmetry in the JL states between the electron and positron
wavefunctions leads to more complicated Lorentz transformation
characteristics for the cyclotron rates for boosts along the magnetic
field, as shall become evident below. The spin-averaged rate should (and
does), of course, reduce to the Sokolov \& Ternov formalism result in
Eq.~(\ref{eq:Gammave_ST}).

\subsection{Reduction to a Latal-Type Form}
 \label{eq:JL_reduction}

For direct comparison with the Sokolov \& Ternov formalism results
above, it is necessary to generalize the cyclotron emission results in
the Appendix of DV78 to treat arbitrary \teq{n\to 0} transitions,
specifically their Eqs.~(A2) and~(A3).  This is a routine S-matrix
calculation in quantum electrodynamics, and essentially parallels the
cyclotron absorption calculations presented in Sec.~II of DV78.  The
reader is therefore referred to DV78 for relevant background and some
definitional material, and here only a detailed outline of the
derivation is provided.  Wherever possible, the same notation as used in
the ST formulation above is adopted.  The starting point is Eq.~(A1) of
DV78, which can be integrated over photon solid angles \teq{d\Omega
=2\pi\, d(\cos\theta )} to yield the spin-dependent cyclotron rate
\begin{equation}
   \Gamma_{n0}^{\zeta}\; =\; 2\pi\, \dover{\fsc c}{\lambar}
      \sum_{\sigma=\perp,\,\parallel} \int d(\cos\theta )\, d\omega\, 
      \dover{dn}{d\omega}\,\Phi_{\sigma}\quad ,
 \label{eq:Gamma_JL}
\end{equation}
where \teq{dn/d\omega = 4 \pi L^3\omega^2/(2\pi \lambar)^3} is the
density of photon states in a box of size \teq{L}; this size establishes
the normalization of the wavefunctions.  Note that \teq{dn/d\omega} is
dimensionless because of the adopted convention of dimensionless photon
energies \teq{\omega}.

A sum over the \teq{\sigma=\perp,\,\parallel} photon polarization states
has been effected, with the standard convention for the labelling of the
photon polarizations being adopted: \teq{\parallel} refers to the state
with the photon's {\it electric} field vector parallel to the plane
containing the magnetic field and the photon's momentum vector, while
\teq{\perp} denotes the photon's electric field vector being normal to
this plane.  Also,
\begin{equation}
   \Phi_{\sigma}\; =\; \dover{1}{\fsc B} \,\dover{1}{cT} 
   \int \dover{L}{\lambar}\,\dover{dq}{2\pi} 
     \int \dover{L}{\lambar}\, \dover{db}{2\pi}\, \Bigl\vert S_{fi} \Bigr\vert^2
 \label{eq:Phi_sigma_def}
\end{equation}
is the relevant integration over the final parallel momentum \teq{q} of
the electron and the coordinate \teq{b} defining the mean location of
the final electron wavefunction perpendicular to {\bf B}.  Here \teq{T}
is the normalizing time for the QED perturbation calculation. The
central contribution to the integrand comes from
\begin{equation}
   S_{fi}\; =\; -i\, \sqrt{ \dover{\fsc\lambar^3}{2\omega L^3}}\, 
       \dover{2\pi}{\sqrt{E_nE_0}} \, \delta (E_n-E_0-\omega )\, V_{fi}\quad ,
 \label{eq:Sfi_def}
\end{equation}
the S-matrix element for cyclotron emission that is analogous to
Eq.~(14) of DV78 that describes cyclotron absorption.  Here
\begin{equation}
   V_{fi}\; =\; \int d^3x\, e^{i\vec{k}.\vec{x}}\,
        u_0^{\dag}(q,\, b,\; \vec{x}) {\cal M}_{\sigma}
        u_n^{(\zeta )}(p,\, a,\; \vec{x})
 \label{eq:Vfi_def}
\end{equation}
is the Fourier transform of the spatial portion of the interaction
amplitude, basically the vertex function of Melrose and Parle (1983),
but with the temporal phase factors already extracted.  The initial
electron wavefunction is \teq{u_n^{(\zeta )}(p,\, a,\; \vec{x})}, and
can possess either \teq{\zeta =\pm 1} spin designation (note that this
spin notation differs from the \teq{s=\pm 1} notation adopted by DV78);
the final electron wavefunction, \teq{u_0^{\dag}(q,\, b,\; \vec{x})}, is
an \teq{n=0}, \teq{\zeta=-1} ground state, so that an explicit spin
label is suppressed.  Expressions for these spatial parts of the Johnson
\& Lippmann eigenfunctions of the Dirac equation are given Eq.~(10) of
DV78; these do not exhibit the same charge conjugation symmetry that the
Sokolov \& Ternov ``transverse polarization'' states do.

In Eq.~(\ref{eq:Vfi_def}), \teq{{\cal M}_{\sigma}} is one of the two
``polarization matrices'' in Eq.~(15) of DV78, representing contracted
products of polarization vectors and the Dirac \teq{\gamma} matrices.
Since the \teq{u_n^{(\zeta )}} and \teq{u_0^{\dag}} functions possess
the dimensions of \teq{(\hbox{length})^{-3/2}}, it follows that
\teq{V_{fi}} and consequently \teq{S_{fi}} and \teq{\Phi_{\sigma}} are
dimensionless. Observe also that the temporal integration over the
wavefunction products has already been performed leading to the
\teq{\delta} function in Eq.~(\ref{eq:Sfi_def}) that expresses energy
conservation.  Finally, note that \teq{a} defines the mean location of
the initial electron wavefunction perpendicular to {\bf B}, i.e. is a
counterpart to \teq{b}.

The squaring of the matrix elements in Eq.~(\ref{eq:Phi_sigma_def}) is
performed using the approach of, for example, Bjorken \& Drell (1964)
for handling the \teq{\delta} function. The result is 
\begin{equation}
   \Phi_{\sigma}\;\to\; \dover{\fsc}{2\omega}\, \dover{2\pi \lambar^3}{L^3}
   \, \dover{\delta (E_n-E_0-\omega)}{E_nE_0}\, \Bigl\vert V_{fi} \Bigr\vert^2
 \label{eq:Phi_sigma_too}
\end{equation}
in parallel with the developments leading to Eq.~(19) of DV78. 
Integrations analogous to those in Eq.~(\ref{eq:Vfi_def}) have been
performed analytically in DV78 for the cyclotron absorption case,
resulting in the appearance of associated Laguerre functions that are
characteristic of magnetized QED calculations involving either free
electrons or electron propagators. Here, since specialization to the
\teq{n'=0} case is made, these functions reduce to simple combinations
of power-laws and exponentials, viz.,
\begin{equation}
   \bigl\vert \Omega_{n0} \bigr\vert^2 
      \; =\; \dover{\kappa^n}{n!}\, e^{-\kappa}\quad ,\quad
      \kappa\; =\; \dover{\omega^2\sin^2\theta}{2B}\;\; .
 \label{eq:Omegasq}
\end{equation}
in the notation of Eq.~(18) of DV78.   As these spatial integrations
embody translational invariance along the field, they yield a
\teq{\delta} function expressing conservation of momentum parallel to
{\bf B}, i.e. establishing \teq{p=q+\omega\cos\theta}, and completing
the conservation laws in Eq.~(\ref{eq:E_p_conserve}).

A simple crossing symmetry manipulation of Eq.~(16) of DV78 facilitates
the evaluation of Eq.~(\ref{eq:Vfi_def}), and also
Eq.~(\ref{eq:Phi_sigma_def}), which can be compared closely with
Eq.~(19) of DV78.  Using the identity in Eq.~(\ref{eq:Omegasq}), and
reverting to the \teq{\mu =\cos\theta} notation, a modest amount of
algebra then yields relatively compact expressions for the
spin-dependent cyclotron emission rate in the Johnson \& Lippmann
formalism:
\begin{eqnarray}
   \Gamma^{\zeta}_{n0} &=& \dover{\fsc c}{4\lambar}\, 
      \int_{-1}^1 d\mu\, \dover{(E_n-E_0)}{E_n(E_0-q\mu)}\; 
      \dover{e^{-\kappa}\, \kappa^{n-1}}{(n-1)!}   \nonumber\\[-5.5pt]
 \label{eq:Gamma_JL_too}\\[-5.5pt]
      &\times & \; (E_n+1)\, (E_0+1)\,
      \Bigl\{ \Lambda^{\zeta}_{\parallel} + \Lambda^{\zeta}_{\perp} \Bigr\}\quad .\nonumber
\end{eqnarray}
The different photon polarization and electron spin combinations yield the
following contributions:
\begin{eqnarray}
   \Lambda^{+1}_{\parallel} &=& \biggl\{ 
        \biggl( \dover{q}{E_0+1} - \dover{p}{E_n+1} \biggr)\, \mu
         - \dover{\omega (1-\mu^2)}{E_n+1}\, \biggr\}^2\nonumber\\[-3.5pt]
 \label{eq:Lambda_par}\\[-3.5pt]
   \Lambda^{-1}_{\parallel} &=& 2nB\; \biggl\{ 
        \biggl( \dover{q}{E_0+1} + \dover{p}{E_n+1} \biggr)\, \dover{\omega (1-\mu^2)}{2nB}
         - \dover{\mu}{E_n+1}\, \biggr\}^2 \nonumber 
\end{eqnarray}
for the \teq{\parallel} polarization, and 
\begin{eqnarray}
   \Lambda^{+1}_{\perp} &=& 
        \biggl( \dover{q}{E_0+1} - \dover{p}{E_n+1} \biggr)^2\nonumber\\[-5.5pt]
 \label{eq:Lambda_perp}\\[-5.5pt]
   \Lambda^{-1}_{\perp} &=& \dover{2nB}{(E_n+1)^2} \nonumber 
\end{eqnarray}
for photons of \teq{\perp} polarization.  It is now almost trivial to
demonstrate that for the \teq{n=1} case, Eq.~(\ref{eq:Gamma_JL_too}) in
concert with Eqs.~(\ref{eq:Lambda_par}) and~(\ref{eq:Lambda_perp})
reduces to Eqs.~(A2) and (A3) of Daugherty \& Ventura (1978), when
\teq{\zeta =+1} and \teq{\zeta =-1}, respectively.  Accordingly,
Eqs.~(\ref{eq:Gamma_JL_too}--\ref{eq:Lambda_perp}) serve as a
generalization of the cyclotron emission rates presented in DV78 to
arbitrary \teq{n\to 0} transitions.

The isolation of the photon polarization contributions in this
development is an attribute that is absent from the exposition in the
Appendix of Daugherty \& Ventura (1978).  Such an isolation expedites
the algebraic developments that reduce this form into expressions that
much more closely resemble those in the Sokolov \& Ternov formulation,
i.e. Eqs.~(\ref{eq:Gamma_ST}) and~(\ref{eq:Gammave_ST}).  The analytic
reduction is routine, but lengthy, involving the expedient use of
several identities.  Among these are
Eqs.~(\ref{eq:E_p_conserve}--\ref{eq:xi_def}), which can be used to
quickly derive
\begin{equation}
   E_n\mu -p \; =\; E_0\mu -q\quad ,\quad
   \xi (E_n-p\mu )\; =\; E_0-q\mu\quad ,
 \label{eq:E_p_identities}
\end{equation}
the latter of which is just a rearrangement of Eq.~(\ref{eq:xi_altern}).  For the
\teq{\zeta=+1} case, it is just a short path to yield the equivalences
\begin{eqnarray}
   \Lambda^{+1}_{\parallel} &=& \biggl\{ -\dover{\omega}{E_n+1}\;
        \dover{E_0+1-q\mu }{E_0+1} \biggr\}^2\quad ,\nonumber\\[-5.5pt]
 \label{eq:Lambda_1_alt}\\[-5.5pt]
   \Lambda^{+1}_{\perp} &=& \biggl\{ -\dover{\omega}{E_n+1}\;
        \dover{(E_0+1)\mu -q}{E_0+1} \biggr\}^2\quad ,\nonumber
\end{eqnarray}
and for \teq{\zeta =-1}, a moderate amount of work to yield 
\begin{equation}
    \Lambda^{-1}_{\parallel} \; =\; 2nB\; \biggl\{ 
        \biggl( 1 + \dover{nB}{E_n+1} \biggr)\, \dover{q}{E_0+1}
         - \dover{p}{E_n+1}\, \biggr\}^2\;\; .
 \label{eq:Lambda_m1_alt}
\end{equation}
With this assembly of tools, and various algebraic manipulations, one is
lead to the following final forms for the spin-dependent cyclotron rates
in the Johnson \& Lippmann formulation:
\begin{equation}
   \Gamma^{\zeta}_{n0}\; =\; \dover{\fsc c}{\lambar}\, \dover{n^n}{(n-1)!}\,
      \dover{B}{E_n} \int_{-1}^1 \dover{e^{-\kappa}\, d\mu}{(E_n-p\mu)^2}
      \dover{(1-\xi)^{n-1}}{(1+\xi)^{n+1}}\; X_{\zeta}\;\; ,
 \label{eq:Gamma_JL_fin}
\end{equation}
where Eq.~(\ref{eq:kappa_def}) gives \teq{\kappa} in terms of \teq{\xi}, and 
the spin-dependent factors in the integrand are
\begin{eqnarray}
  X_{+1} &=& \dover{2nB}{\xi (E_n+1)}\, \biggl\{ 
     \dover{2nB}{E_n-p\mu} - E_n\, (1-\xi ) \biggr\}\quad ,\nonumber\\[-5.5pt]
 \label{eq:X_def}\\[-5.5pt]
  X_{-1} &=& 2 \Bigl( 1 + 2nB - \dover{1}{\xi} \, \Bigr) - X_{+1}\quad .\nonumber
\end{eqnarray}
These are clearly forms resembling the integral that Latal (1986) derived,
though indicating a slightly more complicated differential distribution
\teq{d\Gamma_{n0}^{\zeta}/d\mu} than for the ST case.
The average of the \teq{X_{\zeta}} factors is obviously
\begin{equation}
   \overline{X}\;\equiv\; \dover{1}{2} \Bigl( X_{+1} + X_{-1} \Bigr)
      \; =\; 1 + 2nB - \dover{1}{\xi}\quad ,
 \label{eq:Xbar_def}
\end{equation}
from which it is evident that the spin-averaged cyclotron rate for this
Johnson \& Lippmann formulation is identical to that of the Sokolov \&
Ternov formalism, i.e. Eqs.~(\ref{eq:Gamma_ST})
and~(\ref{eq:Gammave_ST}), as should be the situation.  Such degeneracy
between formalisms also prevails for the power \teq{\Powerave} radiated
when averaged over the initial electron spins.

Note that while the spin-dependent rates are generally different for the
two sets of wavefunctions, in the limit \teq{p\to 0}, it is clear from 
Eq.~(\ref{eq:X_def}) that \teq{X_{\zeta}\to (1-\zeta /\erg_n) \overline{X}} 
so that the two formalisms become degenerate in this limit.  This is a 
consequence of the mathematical identity of the Sokolov \& Ternov and 
Johnson \& Lippmann eigenstates in the special case of \teq{p=0}.

\subsection{Lorentz Boost Characteristics and Analytics}
 \label{sec:JL_Lorentz_analytic}

As with the analysis in Sec.~\ref{sec:STformalism}, it is instructive to
discern the Lorentz transformation behavior of the Johnson \& Lippmann
rates. Such a development closely parallels that in
Sec.~\ref{sec:ST_Lorentz}, again writing the relevant integrals in terms
of variables in the frame in which \teq{p=0}. With the identities
\teq{E_n=\gamma\erg_n} and \teq{1/(E_n-p\mu )= \gamma
(1-\beta\mu_0)/\erg_n}, it can be quickly established that
\begin{equation}
   X_{+1}\; =\; \dover{2nB}{\gamma\erg_n+1}\, \dover{\gamma}{\erg_n}\,
      \Bigl\{ \overline{X} - 2nB\, \dover{\beta\mu_0}{\xi} \Bigr\}\quad ,
 \label{eq:X_1_Lorentz}
\end{equation}
which can be used also for \teq{X_{-1}=2\overline{X}-X_{+1}}.  As for
the cyclotron power considerations of Sec.~\ref{sec:power_ST}, since the
spin-averaged integrand or \teq{d\Gammave/d\mu_0} is even in
\teq{\mu_0}, the term proportional to \teq{\mu_0} in
Eq.~(\ref{eq:X_1_Lorentz}) contributes exactly zero to the integral in
Eq.~(\ref{eq:Gamma_JL_fin}). This simplification clearly also applies to
\teq{X_{-1}}. It then follows that an alternative form for the rate in
Eq.~(\ref{eq:Gamma_JL_fin}) is
\begin{equation}
   \Gamma^{\zeta}_{n0}\; =\;  \dover{\fsc c}{\lambar}\, \dover{n^n}{(n-1)!}\,
      \dover{B}{\gamma\erg_n} \; \biggl\lbrack 1
        - \dover{\zeta (\gamma + \erg_n)}{\erg_n (\gamma\erg_n+1)}
           \, \biggr\rbrack\;  I_n(B)\quad .
 \label{eq:Gamma_JL_final}
\end{equation}
This form conveniently avails itself of the analytic developments of
Sec.~\ref{sec:ST_analytic}, so that no further reduction of integrals is
necessary: the evaluation of \teq{I_n(B)} as a series of Legendre
functions of the second kind in Eq.~(\ref{eq:InB_alt_too}) is
immediately applicable. Such a simplification cannot be applied to the
spin-dependent cyclotron powers, a point that is addressed below.

The Lorentz-transformed expression for the cyclotron rates in
Eq.~(\ref{eq:Gamma_JL_final}) is attractively simple, but is not as
elegant as its equivalent for the Sokolov \& Ternov formulation, as
embodied in Eqs.~(\ref{eq:Gamma_ST}) and~(\ref{eq:Gammave_ST_alt}). 
Here, boosts along the field do not introduce simple time dilation
factors of \teq{1/\gamma}, essentially amounting to a mixing of JL spin
states induced by such transformations. This less than ideal
characteristic is a consequence of the fact that the JL wavefunctions
are eigenstates of a somewhat convoluted spin operator that is not
symmetric under charge conjugation (e.g., see Melrose \& Parle 1983),
nor is it manifestly covariant under boosts along the field.   In
contrast, the ST ``transverse polarization'' wavefunctions are
eigenstates of the component \teq{\mu_z} along {\bf B} of the magnetic
moment operator \teq{\vec{\mu} = \vec{\sigma} - i \vec{\gamma} \times 
\{\vec{p} + e\, \vec{A} \}} (here \teq{\vec{\sigma}} and
\teq{\vec{\gamma}} are matrix vectors for the Dirac algebra), and such
eigenstates exhibit both appropriate symmetry between positrons and
electrons, and simple transformation properties for boosts parallel to
the field.

The sensitivity of the spin-dependent cyclotron rates, and in particular
their Lorentz transformation characteristics, to the choice of spin
eigenfunctions is naturally expected, since the inherently relativistic
effect of spin-orbit coupling is explicitly incorporated in the Dirac
equation in an external field, thereby emphasizing the interpretative
importance of the choice of wavefunctions. The greater elegance of the
results in Sec.~\ref{sec:STformalism} is yet another argument in favor
of usage of Sokolov \& Ternov states for the cyclotron problem, and
therefore for other QED processes where spin-dependent cyclotron rates
are required.

Note that the differential spectrum for spin-dependent considerations
can be routinely obtained in a manner similar to the derivation of
Eq.~(\ref{eq:dGammave_domega_ST}), observing that \teq{\overline{X}} can 
be replaced by \teq{X_{+1}} in Eq.~(\ref{eq:X_1_Lorentz}), or
\teq{X_{-1}=2\overline{X}-X_{+1}}, as desired, and then using the
substitution \teq{\xi = (\erg_n-1/\erg_n)/\omega -1}.

\figureoutpdf{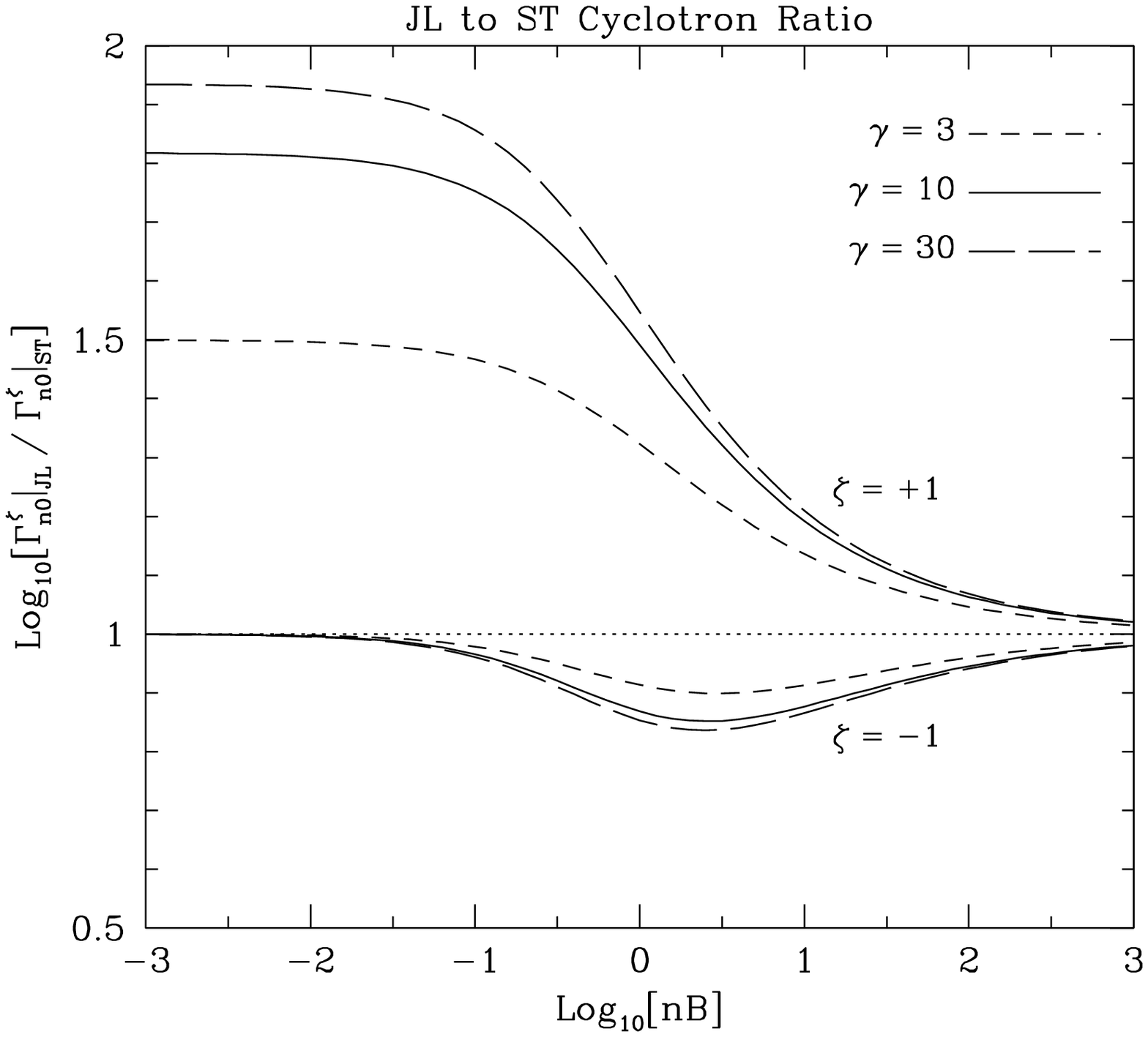}{4.0}{0.0}{-0.2}{
The ratios of the spin-dependent cyclotron decay rates resulting from
Johnson \& Lippmann and Sokolov \& Ternov formalisms, as given in
Eq.~(\ref{eq:Gamma_JL_ST_ratio}).  The rates are plotted as functions of
\teq{nB}, with \teq{B} expressed in units of \teq{B_{\rm cr}=4.413\times
10^{13}}Gauss, for different fixed values of \teq{\gamma}, as labelled,
corresponding to parallel momenta \teq{p=-\gamma\beta\erg_n}.  The upper
three curves correspond to \teq{\zeta =+1}, while the lower three are
for \teq{\zeta =-1}.  The dotted line represents \teq{\gamma=1}, where
the ratio is unity for both \teq{\zeta =\pm 1}.
 \label{fig:cyc_JL_ST_ratio}} 

The ratios of the JL rates in Eq.~(\ref{eq:Gamma_JL_final}) to their ST
counterparts are plotted in Fig.~\ref{fig:cyc_JL_ST_ratio}, being
specifically given by
\begin{equation}
   \dover{\Gamma^{\zeta}_{n0}\Bigl\vert_{\hbox{\sevenrm JL}}}{
      \Gamma^{\zeta}_{n0}\Bigl\vert_{\hbox{\sevenrm ST}}}\; =\;
      \biggl\lbrack 1
        - \dover{\zeta (\gamma + \erg_n)}{\erg_n (\gamma\erg_n+1)}
           \, \biggr\rbrack\; \Biggl/ \; \biggl\lbrack 1
        - \dover{\zeta}{\erg_n} \, \biggr\rbrack\quad .
 \label{eq:Gamma_JL_ST_ratio}
\end{equation}
In general, this ratio is less than unity for \teq{\zeta =-1} and
greater than unity for \teq{\zeta =1}; it is also clearly a
monotonically increasing function of \teq{\gamma} for \teq{\zeta =+1},
and decreases monotonically with \teq{\gamma} for \teq{\zeta =-1}. This
ratio approaches unity for either \teq{\gamma\to 1} (dotted line) or for
\teq{2nB\gg 1}, and in the case of \teq{\zeta =-1} also for \teq{2nB\ll
1}.  In the limit \teq{2nB\ll 1}, the ratio approaches
\teq{2\gamma/(\gamma+1)}  when \teq{\zeta=1}, the maximum departure from
unity in this case. In the \teq{2nB\sim 1} domain, as
\teq{\gamma\to\infty}, for \teq{\zeta =-1} the minimum in the ratio is
\teq{(4+2\sqrt{2})/(4+3\sqrt{2})\approx 0.83} and occurs for
\teq{nB=1+\sqrt{2}}.  

The significant deviations from unity underline the importance of the
choice of wavefunctions when retaining spin dependence.  This becomes a
profound issue for resonant Compton scattering problems (e.g. see
Graziani 1993), where, as discussed above, the spin-dependent cyclotron
lifetime is required to render the intermediate electron states
metastable and thereby truncate the divergent resonance at the cyclotron
frequency. It is also pertinent to handling resonances in magnetic
Coulomb scattering (e.g. Langer 1981), two-photon pair creation in
strong fields (e.g. Kozlenkov \& Mitrofanov 1986), magnetic photon
splitting above pair creation threshold (discussed briefly in Baring
2000), and other general external field problems involving electrons
either in free states or as propagators (Graziani, Harding \& Sina,
1995).

To conclude this section, a brief exposition of the spin-dependent
powers is offered.  The spin-averaged power is, of course, that in
Eq.~(\ref{eq:Powerave_ST_alt}).  To retain spin dependence in the JL
formalism, one proceeds by weighting the integrand of
Eq.~(\ref{eq:Gamma_JL_fin}) with a \teq{\omega =2nB/(1+\xi )/(E_n-p\mu
)} factor. Guided by the analysis in Sec.~\ref{sec:power_ST}, terms that
are odd in \teq{\mu_0} contribute zero, and the by now routine
manipulations yield the form
\begin{eqnarray}
   P_{n0}^{\zeta}  &=& \dover{\fsc m_ec^3}{\lambar}\, \dover{n^{n+1}}{(n-1)!}\,
      \dover{B^2}{\erg_n^2} \; \nonumber\\[-3.5pt]
 \label{eq:Power_JL_spins}\\[-3.5pt]
    &\times & \hskip -5pt \Biggl\{ \biggl\lbrack 1
        - \dover{\zeta (\gamma + \erg_n)}{\erg_n (\gamma\erg_n+1)}
           \, \biggr\rbrack\, K_n(B) + \zeta\; \dover{2\gamma\beta^2\, (nB)^3}{\erg_n^3
          (\gamma\erg_n+1)}\, L_n(B) \Biggr\} \, ,\nonumber
\end{eqnarray}
where \teq{K_n(B)} is the integral in Eq.~(\ref{eq:KnB_def}),
or equivalently, the series in Eq.~(\ref{eq:KnB_alt_too}), and
the \teq{L_n(B)} term emerges from a \teq{\mu_0^2} term in the
integrand resulting from the product \teq{\omega X_{\zeta}},
and corresponds to
\begin{equation}
   L_n(B)\; =\; \sum_{k=0}^{\infty} \dover{(-n)^k}{k!}\, 
             \dover{J_{n+k}(z_n)}{n+k+1}\quad .
 \label{eq:LnB_def}
\end{equation}
Eq.~(\ref{eq:Power_JL_spins}) is clearly not as elegant as the
corresponding ST result, namely \teq{P_{n0}^{\zeta} =
(1-\zeta/\erg_n)\Powerave}, but does reduce to this simpler form in the
limit \teq{\beta\to 0}, as expected. Furthermore, averaging
Eq.~(\ref{eq:Power_JL_spins}) over spins \teq{\zeta =\pm 1} then yields
exactly \teq{\Powerave}, so that degeneracy between the JL and ST
formalisms is realized when spin information is not explicitly retained.
Note that asymptotic limits of \teq{P_{n0}^{\zeta}} for small and large 
\teq{nB} can be readily obtained but are not particularly elucidating for 
the purposes of this exposition.

\section{CONCLUSION}
 \label{sec:conclusion}

This paper presents useful analytic developments of spin-dependent
cyclotron decay rates/widths for general \teq{n\to 0} transitions
between Landau states, expressing these in terms of various series of
Legendre functions \teq{Q_{n}(z)} of the second kind. The rates and the
radiated cyclotron powers are derived using two popular eigenstate
formalisms appropriate for the Dirac equation in a uniform magnetic
field, namely that due to Sokolov \& Ternov (1968), and that due to
Johnson \& Lippmann (1949). The resulting expressions are
Eq.~(\ref{eq:Gammave_ST_red}) for the cyclotron rate and
Eq.~(\ref{eq:Powerave_ST_alt}) in conjunction with
Eq.~(\ref{eq:KnB_alt_too}) for the Sokolov \& Ternov formulation, and
Eq.~(\ref{eq:Gamma_JL_final}) for the rate and
Eq.~(\ref{eq:Power_JL_spins}) for the power for the Johnson \& Lippmann
case. These compact forms can be expediently used in astrophysical
models, since they generally amount to rapidly convergent series of
elementary functions.  These spin-dependent results are particularly
important for addressing resonant cyclotronic divergences that appear in
various higher-order QED mechanisms in external magnetic fields.  The
Johnson \& Lippmann developments extend the work of Daugherty \& Ventura
(1978) to arbitrary \teq{n\to 0} transitions, while the series
development permits a more accurate determination of \teq{n B\gg 1}
asymptotics of the Sokolov \& Ternov rates than that offered in Latal
(1986).

In addition, the Lorentz transformation characteristics of the rates are
derived for boosts along the magnetic field, yielding a simple time
dilation multiplicative factor in the ST case, but a slightly more
complicated and less ideal dependence on boost Lorentz factor for the JL
analysis, where spin-state mixing is implied by such boosts.  Such
comparitive complexity of the JL formalism is more pronounced for the
radiated power.  This comparison again underlines the superior
attributes of Sokolov \& Ternov eigenstates for use in evaluating rates
and cross sections of magnetized processes in quantum electrodynamics.

\vskip 40pt
\acknowledgments 
We thank the referee, Carlo Graziani, for incisive comments that were
helpful to the polishing of the manuscript.  We also thank Joe Daugherty
for the provision of materials pertaining to the Johnson \& Lippmann
formalism, and PLG acknowledges the assistance of Stefan Coltisor during
the early stages of the development of the decay rates using the Johnson
\& Lippmann wavefunctions.  In performing this research, we are grateful
for the generous support of the NASA Astrophysics Theory Program, the
Research Corporation (grant CC5813), and the National Science Foundation
(REU and grant AST-0307365). 

\appendix
\section{Properties Involving the Associated 
Legendre Functions of the Second Kind, \teq{J_{\nu}(z)} }
 \label{sec:appendix}

This Appendix provides several identities germane to the developments 
in the text, principally in relation to the Legendre functions of the 
{\it second kind}, \teq{Q_{\nu}(z)}, which form the basis of the 
series expansions for the cyclotron rates.  They can be defined in terms of 
hypergeometric functions (e.g. see Eq.~8.820.2 of Gradshteyn 
\& Ryzhik 1980), or alternatively the integral representation in 
Eq.~(\ref{eq:Qnu_def}).  For integer indices \teq{\nu =n}, the specialization
of interest here, they are simple combinations of logarithmic and
polynomial functions (e.g. see Abramowitz \& Stegun 1965), that lead 
to similar simplicity for the \teq{J_{\nu} (z)} using Eq.~(\ref{eq:Jnu_ident}).
Observe that \teq{Q_{\nu}(z)\equiv Q_{\nu}^0(z)}, where the
associated Legendre functions \teq{Q_{\nu}^{\mu}(z)} are defined in 8.703 of
Gradshteyn \& Ryzhik (1980).  Note also the alternative identities
\begin{equation}
   J_{\nu} (z) \; =\; -(\nu +1)\,\sqrt{z^2-1}\; Q^{-1}_{\nu}(z)
      \; =\; (\nu +1)\int_z^{\infty} Q_{\nu}(t)\, dt\quad , \quad z\; >\; 1\quad ,
 \label{eq:J_nu_ident}
\end{equation}
embodied in the results 8.734.3 and 8.752.5 in Gradshteyn 
\& Ryzhik (1980).  Explicit functional forms for some of the \teq{Q_n(z)}
are given in Abramowitz \& Stegun (1965) and Eq.~8.827 of Gradshteyn 
\& Ryzhik (1980), from which the following identities for the first few 
\teq{J_n(z)} can be formed:
\begin{eqnarray}
   J_1(z) &=& z - \dover{z^2-1}{2}
                    \, \log\Bigl\vert \dover{z + 1}{z - 1} \Bigr\vert\quad ,\nonumber\\[2pt]
   J_2(z) &=& -1 + \dover{3z^2}{2} - \dover{3z\, (z^2-1)}{4}
                    \, \log\Bigl\vert \dover{z + 1}{z - 1} \Bigr\vert\quad ,
  \label{eq:J_n_ident}\\[2pt]
   J_3(z) &=& -\dover{13z}{6} + \dover{5z^3}{2} - \dover{(5z^2-1)\, (z^2-1)}{4}
                   \, \log\Bigl\vert \dover{z + 1}{z - 1} \Bigr\vert\quad .\nonumber
\end{eqnarray}
An effective way to derive expressions for the \teq{J_n(z)} in terms
of elementary functions, at least for low \teq{n}, is to use the
Rodrigues formula in 8.836.1 of Gradshteyn \& Ryzhik (1980), and then
invoke Eq.~(\ref{eq:Jnu_ident}). An alternative approach is to use
identities 8.831.2 and 8.831.3 of Gradshteyn \& Ryzhik (1980), to arrive
at a relation between \teq{Q_n(z)} and a sum of Legendre polynomials
\teq{P_s(z)} and a logarithmic term.  This quickly yields
\begin{eqnarray}
  J_n(z) &=& Q_{n-1}(z) - z\, Q_n(z)\nonumber\\[-5.5pt]
 \label{eq:J_n_series}\\[-5.5pt]
              &=& \dover{1}{2}\,\Bigl[ P_{n-1}(z) - z\, P_n(z)\Bigr]
                       \, \log\Bigl\vert \dover{z + 1}{z - 1} \Bigr\vert
              - \sum_{k=1}^{n-1} \dover{1}{k}\, P_{k-1}(z)\, P_{n-1-k}(z)
                    + z \sum_{k=1}^{n} \dover{1}{k}\, P_{k-1}(z)\, P_{n-k}(z)\quad ,\nonumber
\end{eqnarray}
a form that clearly displays the combined logarithmic and polynomial
character of these functions, and which is useful for asymptotic
results, explored just below. 

There are various possible approaches to
numerical calculations of the \teq{J_n} functions for integer indices. 
Options include direct use of Eq.~(\ref{eq:J_n_series}), combined with
an efficient algorithm for computing the Legendre polynomials for
arguments \teq{z>1}, such as invoking recurrence relations (e.g. see
Eq.~8.5.3 of Abramowitz \& Stegun 1965), which are numerically accurate
for upward recurrence in the polynomial index \teq{n}. Alternatively,
direct recurrence of the \teq{Q_n(z)} functions, which satisfy the same
relations as the \teq{P_n(z)}, is viable, though this is only stable for
downward iterations in \teq{n} when \teq{z>1}.  This approach therefore
requires evaluation of the \teq{P_n(z)} for two particular initial
values \teq{n}, perhaps via the hypergeometric function series using the
identity in Eq.~8.1.5 of Abramowitz \& Stegun (1965). However, probably
the most expedient numerical technique is to use the recurrence relation
for the \teq{J_n(z)} functions directly:
\begin{equation}
   n(n+1)\, J_{n+1}(z) - n(2n+1)\, z\, J_{n}(z) + (n^2-1)\, J_{n-1}(z)\; =\; 0\quad .
 \label{eq:Jrecurrence}
\end{equation}
This identity can be deduced after a modicum of algebra from Eq.~8.5.3
of Abramowitz \& Stegun (1965).  Note that strong cancellation often
arises between the last two terms of this equation, so that for large
\teq{n} and \teq{z >1}, the values of \teq{J_n(z)} decline exponentially
with \teq{n} according to Eq.~(\ref{eq:J_n_lownB}). Starting with
initial values prescribed by Eq.~(\ref{eq:J_n_ident}), this technique is
stable to upward recurrence in \teq{n} for any \teq{z>1}, the range of
arguments appropriate for the cyclotron problem.  This quickly yields
suitably precise values (numerically tested for \teq{n\leq 100}) for a
range of indices required in the series summations that appear in the
cyclotron rates and powers, for example in
Eqs.~(\ref{eq:Gammave_ST_red}) and~(\ref{eq:KnB_alt_too}). The use of
this recurrence becomes superfluous typically for \teq{n>30}, since such
indices are generally required only in the classical cyclotron limit
when \teq{z_n\gg 1}, and the asymptotic limit in
Eq.~(\ref{eq:J_n_lownB}) is then the preferred computational tool.

Now focusing on asymptotics issues, two limiting values of \teq{J_n(z)}
are required for the analysis in Sec.~\ref{sec:ST_analytic}.  The
\teq{z\gg 1} limit is routinely derived, and is listed in
Eq.~(\ref{eq:J_n_lownB}).  Since \teq{z\geq 1}, the other limiting
domain is \teq{z\to 1}.  This is a somewhat more involved case. By
inspection of Eq.~(\ref{eq:J_n_series}), as \teq{z\to 1}, the
logarithmic terms approach zero since \teq{P_n(z)\to 1 + n(n+1) (z-1)/2
+ O(z-1)^2}. The Legendre polynomials thus simplify to approximately
unity, to leading order, and it follows that in the neighborhood of
\teq{z=1},
\begin{equation}
   J_n(z)\; =\; \dover{1}{n} -\dover{n+1}{2}\, (z-1)
                \, \log\Bigl\vert \dover{2}{z - 1} \Bigr\vert + O(z-1)
 \label{eq:J_n_Taylor_zeq1}
\end{equation}
defines the leading order terms to the Taylor series expansion.  Insertion in
Eq.~(\ref{eq:InB_alt_too}) then yields
\begin{equation}
   I_n(B) \; \approx\; \sum_{k=0}^{\infty} \dover{(-n)^k}{k!}\, 
              \biggl\{ \dover{1}{n+k} -\dover{n+k+1}{2}\, (z_n-1)
                \, \log\Bigl\vert \dover{2}{z_n - 1} \Bigr\vert\biggr\} 
              \; =\; \dover{\gamma (n,\, n)}{n^n}
                        - e^{-n}\, \dover{(z_n-1)}{2}\, \log\Bigl\vert 
                        \dover{2}{z_n - 1} \Bigr\vert \quad ,
 \label{eq:InB_Taylor_series}
\end{equation}
correct to order \teq{O([z-1]\, \log\vert z-1\vert\, )}.
Here, \teq{\gamma (n,\, x)} is the incomplete Gamma function, and 
can be introduced using the series identity 8.354.1 of Gradshteyn 
\& Ryzhik (1980). 
 
Another identity required for the development of
Eq.~(\ref{eq:Gammave_highnB_too}) pertains to the evaluation of
\teq{\gamma (n,\, n)} for large \teq{n}.  For low values of \teq{n}, say
\teq{n\lesssim 20}, one can use 8.352.1 of Gradshteyn \& Ryzhik (1980),
namely
\begin{equation}
   \dover{\gamma (n,\, n)}{\Gamma (n)}
      \;\equiv\; 1 - \dover{\Gamma (n,\, n)}{\Gamma (n)}
      \; =\; 1 - e^{-n} \sum_{k=0}^n \dover{n^k}{k!}\quad ,
 \label{eq:gamma_inc_ser}
\end{equation}
as an efficient series evaluation for either the \teq{\gamma (n,\, n)}
or \teq{\Gamma (n,\, n)} incomplete Gamma functions.  For larger
\teq{n}, one can appeal to the identity \teq{\Gamma (n+1,\, n)=n\,
\Gamma(n, \, n) + n^n e^{-n}}, together with Stirling's expansion
\teq{\Gamma (n) \approx e^{-n} n^{n-1/2} \sqrt{2\pi}} and identity
6.5.25 of Abramowitz \& Stegun (1965), namely \teq{\Gamma (n+1,\, n)
\approx e^{-n} n^n \sqrt{n\pi /2}}, to establish the results 
\begin{equation}
   \lim_{n\to\infty} \dover{\gamma (n,\, n)}{\Gamma (n)} \; =\; \dover{1}{2}
   \; =\; \lim_{n\to\infty} \dover{\Gamma (n,\, n)}{\Gamma (n)}\quad .
 \label{eq:gamma_inc_limit}
\end{equation}
Retaining the next order contributions to \teq{\gamma (n,\, n)} and
\teq{\Gamma (n,\, n)} then yields corrections of approximately
\teq{ \pm 1/(3\sqrt{2n\pi}\, )} to the two ratios in Eq.~(\ref{eq:gamma_inc_limit}).

\end{document}